\documentclass[aps,prb,twocolumn,floats,showpacs,superscriptaddress]{revtex4-1}
\usepackage{graphicx,epsfig}
\usepackage{times}
\usepackage{graphics,dcolumn,bm,float}
\usepackage[caption=false]{subfig}
\usepackage{amssymb,amsmath,rotate,color}
\usepackage[title,titletoc,toc]{appendix}
\usepackage[pagebackref=false,colorlinks,linkcolor=blue,citecolor=blue,urlcolor=magenta]{hyperref}
\usepackage{subfig}

\begin{document}
\unitlength 1 cm
\newcommand{\be}{\begin{equation}}
\newcommand{\ee}{\end{equation}}
\newcommand{\bearr}{\begin{eqnarray}}
\newcommand{\eearr}{\end{eqnarray}}
\newcommand{\nn}{\nonumber}
\newcommand{\la}{\langle}
\newcommand{\ra}{\rangle}
\newcommand{\cd}{c^\dagger}
\newcommand{\vd}{v^\dagger}
\newcommand{\ad}{a^\dagger}
\newcommand{\bd}{b^\dagger}
\newcommand{\tk}{{\tilde{k}}}
\newcommand{\tp}{{\tilde{p}}}
\newcommand{\tq}{{\tilde{q}}}
\newcommand{\eps}{\varepsilon}
\newcommand{\vk}{\vec k}
\newcommand{\vp}{\vec p}
\newcommand{\vq}{\vec q}
\newcommand{\vkp}{\vec {k'}}
\newcommand{\vpp}{\vec {p'}}
\newcommand{\vqp}{\vec {q'}}
\newcommand{\bk}{{\bf k}}
\newcommand{\bp}{{\bf p}}
\newcommand{\bq}{{\bf q}}
\newcommand{\br}{{\bf r}}
\newcommand{\bR}{{\bf R}}
\newcommand{\up}{\uparrow}
\newcommand{\down}{\downarrow}
\newcommand{\fns}{\footnotesize}
\newcommand{\ns}{\normalsize}
\newcommand{\cdag}{c^{\dagger}}

\newcommand{\sx}{\sigma^x}
\newcommand{\sy}{\sigma^y}
\newcommand{\sz}{\sigma^z}

\title{Quantitative measure for the spin-charge separation in two dimensional Hubbard model}

\author{A. Youssefi}
\affiliation{Department of Physics, Sharif University of Technology, Tehran 11155-9161, Iran}
\affiliation{Department of electrical engineering, Sharif University of Technology, Tehran, Iran}

\author{M. Ansari Fard}
\affiliation{Department of Physics, Sharif University of Technology, Tehran 11155-9161, Iran}

\author{S. A. Jafari}
\email{akbar.jafari@gmail.com}
\affiliation{Department of Physics, Sharif University of Technology, Tehran 11155-9161, Iran}
\affiliation{Center of excellence for Complex Systems and Condensed Matter (CSCM), Sharif University of Technology, Tehran 1458889694, Iran}

\begin{abstract}
We introduce a quantitative measure of spin-charge separation, $\zeta(t)$ which is based on the difference
between the fluctuations with respect to background of the spin and charge profiles at any time $t$ and
is suitable for studying the non-equilibrium dynamics of excitations in strongly correlated systems.
This quantity is not only a direct measure of the spin-charge separation in strongly correlated systems,
but its long time behaviour can further serve as a possible order parameter for the interaction
induced (Mott) insulating state. Within the numerically exact diagonzalization we calculate this
quantity for the two dimensional Hubbard model away from Half filling. Our quantitative measure
in chain, ladder and two-dimensional geometries gives the same order of magnitude for the quantity of 
spin-charge separation. Furthermore from the temporal behaviour of $\zeta(t)$ a threshold time
can be identified that provides clues onto the breakdown of underlying Mott insulating phase.
\end{abstract}

\maketitle

\section{Introduction}
Spin and charge are part of the identity of an electron as a fundamental particle
in vacuum. But in condensed matter where a bunch of electrons come together,
if the spatial dimension is restricted to one dimension (1D), these two quantum characteristics
of electrons tend to behave as if they are distinct entities which is referred to
as the spin-charge separation (SCS).
In one dimensional conductors the spin and charge move with two different
velocities that is determined by Coulomb interaction strength~\cite{Tomonaga, Luttinger,Voit}.
This makes the interacting electron liquids in 1D quite different from those in
three-dimensional Fermi liquids. While the excitations of three dimensional Fermi liquids are 
electron and hole-like quasiparticles, in the excitation spectrum of 1D liquids
known as Tomonaga-Luttinger liquids there are no such quasiparticles,
but instead there are collective modes that carry spin-only or charge-only~\cite{Giamarchi}.
This remarkable phenomenon has been experimentally observed in 1D 
GaAs/AlGaAs heterostructures where spin and charge modes with different velocities
were identified~\cite{West}. Characteristics of tunneling into 1D spin-charge
separated systems are also observed~\cite{Schofield}. Also signatures of spin-charge
separation have been observed in variety of systems including 
carbon nano-tubes~\cite{McEuen}. The photoemission spectra of Au chains on Si(111) surface 
show power-laws predicted by Tomonaga-Luttinger theory~\cite{Baer,Nagaosa}.
Electromagnetic response of class of organic salts~\cite{Schwartz} is also consistent with
the picture of spin-charge separation.
For insulating 1D materials where charge degrees of freedom
are localized by strong Coulomb interactions, the spin degrees of freedom retain
their kinetic energy and can roam about. Angular resolved photoemission data in 
1D copper-oxide chain material SrCuO$_2$ backed by exact diagonalization calculations
of 1D t-J model support the picture of spin and charge separation in 
1D insulators~\cite{Kim} as well. This separation gives rise to large optical
nonlinearity in such Mott insulators~\cite{Tohyama}.
Theoretical understanding of the separation of spin and charge in Mott insulators is based on
the work of Ogata and Shiba~\cite{OgataShiba} who used the exact Bethe ansatz
solution of Lieb and Wu~\cite{LiebWu} to prove exactly that the ground state of the 1D
Hubbard model in the limit of very large Coulomb repulsion factorizes into 
a Slater determinants of spin-less fermions and a bosonic wave function (even with respect
to particle exchange) that is the ground state of a related spin chain Hamiltonian. 
Implications of spin-charge separation in the dielectric response of a 
1D Mott insulator compound Sr$_2$CuO$_3$ were also experimentally investigated~\cite{Uchida}.

Therefore the low-energy part of the spectrum of excitations in 1D systems are exhausted by collective
modes of charge and spin. In 1D conductors the charge mode is gapless,
while in the 1D insulators the charge mode is gapped by strong Coulomb interactions.
While the quasi particle excitations in three dimensional Fermi liquids are electron and hole-like,
and in 1D are collective spin and charge modes, understanding the nature of excitations 
in two-dimensional (2D) interacting systems remains an unsettled quest.
Many unusual properties of strongly correlated two-dimensional cuprate superconductors
are ascribed to some form of non-Fermi liquid behaviour~\cite{Anderson2DLuttinger}. 
Anderson emphasizes that the spin-charge separation is the key to understanding
the physics of high temperature cuprates~\cite{AndersonPhysC}.
Standard approaches to 
non-Fermi liquid states are based on auxiliary particle methods~\cite{RMPslave} 
which build on the {\em assumption} of separate particles carrying spin and charge
of the physical electron. Dynamics of a single-hole in a frustrated 2D quantum
magnet supports the picture of spin-charge separation~\cite{Poilblanc}.
Quantum Monte Carlo study of the t-J model also supports the SCS in 2D~\cite{TKLee}.
Furthermore cluster perturbation theory study of spectral properties of 
1D and 2D Hubbard model also supports the SCS in two-dimensional Hubbard model~\cite{Senechal}.

Since in 2D there is no exact solution akin to Leib-Wu solution of the 1D Hubbard model, 
although other methods based on quantum Monte carlo may also be applicable~\cite{Hanke,simone},
the unbiased method of choice to study the exact dynamics of excitations in 2D system 
is the numerically exact diagonalization~\cite{JafariED} and related technique of density
matrix renormalization group~\cite{white,rmpdmrg}. This approach was already
taken by Jagla, Hallberg, and Balseiro in 1993~\cite{Balseiro}. Using very small
clusters affordable in 1993, and with qualitative pictures they concluded that
spin and charge do separate in 1D, while in 2D there is no sign of spin-charge
separation. The 1D aspect of this work was revisited by Kollath and coworkers using time-dependent
density-matrix renormalization group to study the real time dynamics of the 
Hubbard model in a 1D chain who confirmed the spin-charge separation beyond the
low-energy theory of Tomonaga-Luttinger liquid~\cite{Kollath}. They applied 
the model to describe cold Fermi gases in a harmonic trap.

In this work we revisit this problem and introduce a direct and {\em quantitative} 
measure for the separation of spin and charge degrees of freedom. 
Our time-dependent quantity, $\zeta(t)$, is defined as the difference in the
profiles of spin and charge density averaged over the entire space
satisfies the following properties:
(1) It is zero at $t=0$ when a test particle is added to the ground state of the
system, and increases with time when we have spin-charge separation. (2) For $U=0$ limit
of the Hubbard model where the ground state is a Slater determinant it remains
zero at all later times $t$. (3) The $\lim_{t\to\infty} \zeta(t)$ is zero for
conducting phase, and non-zero for Mott insulating phase, and therefore $\zeta(\infty)$
can serve as an {Mott insulating order parameter} obtained from non-equilibrium dynamics.
Therefore this quantity contains information about the
nature of charge localization in the Mott phase. Equipped with this quantity, 
we undertake exact diagonalization investigation of one and two-dimensional Hubbard model 
and we find quite surprisingly that within our measure, the order of magnitude of 
the spin-charge separation quantity in 2D is the same as the one in 1D. Our non-equilibrium study
of the excitation dynamics further shows that in the Mott insulating phase, the
charge carriers are localized in long time-scales, while in the short time-scales
the charge density fluctuates spatially. The cross-over between these two regimes
can provides clues to the breakdown of Mott insulating phase.

\section{Model and method}
We consider the Hubbard model given by 
\begin{equation}
   \hat H= - \sum_{\la i,j\ra,\sigma} \hat c^{\dagger}_{i\sigma} \hat c_{j\sigma} + 
   + U \sum^{L}_{i=1} \hat n_{i\uparrow} \hat n_{i\downarrow},
\end{equation}
where $c^\dagger_{i\sigma}$ creates an electron at site $i$ with spin $\sigma$, the hopping
amplitude has been set as the unit for the energy, the Hubbard $U$ is the on-site
interaction strength, $\hat n_{i\sigma}=c^\dagger_{i\sigma}c_{i\sigma}$ is the fermion occupation number
and $L$ is the number of lattice sites. 
The site label $i$ can refer to any lattice.

In this work we will consider the real-time
evolution of an electron added to the ground state of the above Hubbard model in 1D, ladder and 2D square lattice
and various fillings. The lattice constant in this work are assumed to 
be the unit of length. 
In our simulation we start with the ground state of a system with $ N=N_\up+N_\down$ electrons and
then add a new electron at momentum $\la k_x\ra=\pi/2$. 
After adding the new electron the total number of electrons becomes $N+1$ 
which defines the filling fraction as $n=(N+1)/L$.

\begin{figure}[t]
\includegraphics[width = 0.9\columnwidth]{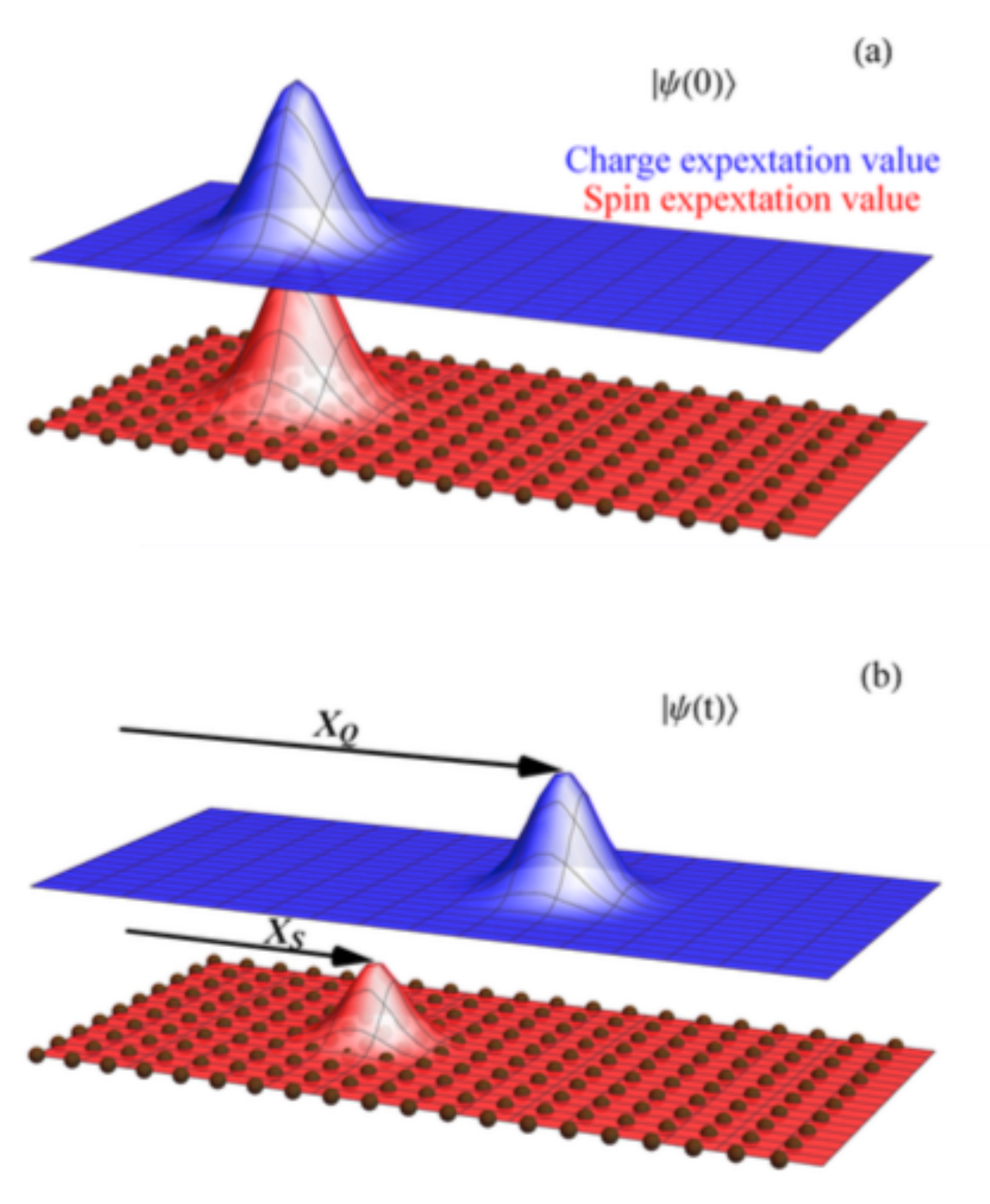}
\caption{(Color online) Schematic representation of separation of charge (blue)
and spin (red). (a) At initial time $t=0$ spin and charge are concentrated 
around the same region. (b) After the wave function has evolved by a large $U$ Hubbard
Hamiltonian, the spin and charge densities are piled in different regions of space
and propagate with different velocities. Quantities $X_Q$ and $X_S$ are 
"center of mass" of the charge and spin as defined in Eq.~\eqref{xqxs.eqn}.
}
\end{figure}

 In the two-dimensional case
$k_y=0$, meaning that the initial wave-packet has no spatial dispersion along the $y$ 
direction. We calculate the ground state wave function with the standard
Lanczos algorithm. Then at time $t=0$ we add an electron in the wave-packet
with a non-zero kinetic energy and denote the resulting state with $|\psi(t=0)\ra$. 
The kinetic energy of the added wave-packet moves it
forward. At a later time $t$ the wave function evolves to $|\psi(t)\ra$
according to standard quantum evolution formula~\cite{Thijssen},
\be
   |\psi(t+dt)\ra \approx \left[1-i\hat H dt-\hat H^2(dt)^2/2\right]|\psi(t)\ra,
   \label{ev1.eqn}
\ee
where we have set $\hbar=1$.
At every step of time-evolution we normalize the wave function
to ensure the stable propagation up to longer times. 
We have explicitly cross-checked the results of above evolution
algorithm against the brute force dynamics,
\be
   |\psi(t)\ra =\sum_{n} e^{-iE_nt}|\psi(t=0)\ra,
\ee
for small few-particle Hilbert spaces where all excited states energies
$E_n$ are numerically accessible by exact diagonalization of the Hamiltonian
matrix. Typical values of $dt$ in our natural units (where $\hbar$ and hopping
amplitude are set to unity) are $10^{-4}$.
We have furthermore checked that the quantum evolution given by 
Eq.~\eqref{ev1.eqn} is stable with respect to variation in $dt$ across two 
orders of magnitude $dt=10^{-3}-10^{-5}$. 

The spin and charge density operators $\hat S_i$ and $\hat Q_i$ at every lattice 
site can be defined as,
\begin{equation}
\hat S_i = (n_{i\uparrow} - n_{i\downarrow}),~~~~~~
\hat Q_i = (n_{i\uparrow} + n_{i\downarrow}).
\end{equation}
Notice that we have deliberately dropped the factor of $\dfrac{1}{2}$ in definition of spin
density. This will be clear when we define our quantitative measure of the spin-charge separation. 
At any later time $t$ where the wave function has evolved to $|\psi(t)\ra$, we can
obtain the expectation values of the above operators to define
$\la {S}_i(t) \ra= \la \psi(t)|\hat S_i|\psi(t)\ra $ and $\la Q_i(t)\ra = \la \psi(t)|\hat Q_i|\psi(t)\ra $.
The spatial profile of $\la S_i(t)\ra$ and $\la Q_i(t)\ra$ as a function of space variable $i$
coincide at $t=0$. At later times the spatial profiles of charge and spin start to 
differ from each other when large enough Hubbard $U$ is used to generate the dynamics
of an added electron. 
As pointed out by Jagla {\em et al}~\cite{Balseiro} the
SCS can be qualitatively seen in 1D as the separation of the peaks of the spatial
profiles of the above two functions. The system sizes considered in 1993 in Ref.~\cite{Balseiro} 
and the qualitative assessment lead them to conclude that within their
exact diagonalization study they do not find spin-charge separation in 2D square lattice.
In this work we consider much larger system sizes and furthermore define 
the quantity of spin-charge separation as follows:

The spatially averaged background for the spin and charge densities before adding the new electron 
is given by $ S_{\rm bg}=0$ and $ Q_{\rm bg}=N/L$. The spatial 
fluctuations defined by $\delta S_i=\la S_i(t)\ra- S_{\rm bg}$
and $\delta Q_i=\la Q_i(t)\ra - Q_{\rm bg}$
can be used to define,
\be
X_Q(t) = \sum_{i=1}^L \delta Q_i(t)~ x_i,~~~
X_S(t) = \sum_{i=1}^L \delta S_i(t)~ x_i,
\label{xqxs.eqn}
\ee
which is spatial average of charge and spin densities and can be viewed as
the worldline of the center of mass of charge and spin. The standard deviations from
the above averages 
can also be defined to assign an "error bar" (due to quantum effects) to each of the 
worldlines. The above quantities represent the charge and magnetic
polarization of the medium. 
Now we are ready to define the quantity of spin-charge separation as,
\be
\zeta^2(t)= \dfrac{1}{L} \sum^{L}_{i=1} (\delta {Q}_i - \delta{S}_i)^2,
\label{zeta.eqn}
\ee
which is nothing but the spatial average of the difference in the {\em fluctuations}  of
charge and spin density at every instant $t$ of time.
At $t=0$ charge and spin have identical profiles therefore one always has $\zeta(t=0)=0$. Moreover at $U=0$ where the ground state of system
is a simple slater determinant this quantity remains zero for all later times. 
We show that the quantity $\zeta(t)$ has a great deal of information not only on the
separation of spin and charge, but also on the localization of charge in the Mott
insulator and its associated time scales. 
Indeed the $\zeta(t)$ being some sort of fluctuation will contain -- within the 
general fluctuation-dissipation theorem -- information about the response of
system to external perturbation: As we will discuss in this work, from the
short time behaviour of $\zeta(t)$, one can infer information on the breakdown
of a Mott insulating state in response to an strong applied voltage. 

Therefore the function $\zeta(t)$: (i) Quantifies the spin-charge separation
and hence provides a measure to compare the amount of separation in various settings,
e.g. between one and two dimensions. (ii) This measure understands the difference
between the Mott insulating and conducting phases. (iii) This measure has ideas
about the non-equilibrium charge (and spin) dynamics and essential time (and hence
energy) scales at which a Mott insulator behaves as an insulator. The later means
that instead of applying a high enough voltage to study the breakdown of Mott
insulator as has been previously done within the exact diagonalization method
by Oka and collaborators~\cite{Oka},
we consider the evolution of quantity $\zeta$ with time. A Mott insulating
behaviour sets in, only after a threshold time which can be associated via
the uncertainity principle with a breakdown. 

\begin{figure}[tb]
   \includegraphics[width=0.23\textwidth]{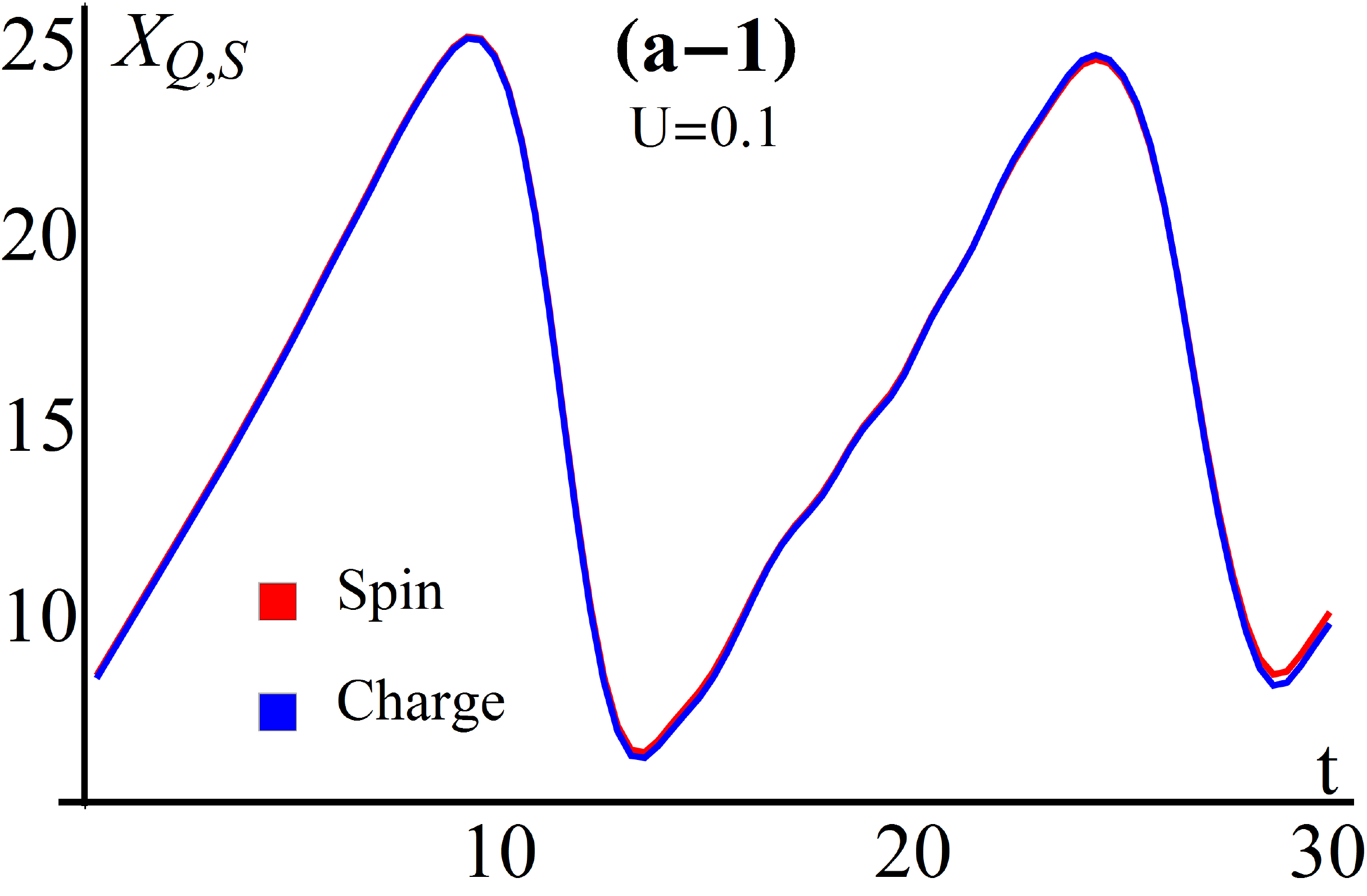}\hspace{3mm}
  \includegraphics[width=0.23\textwidth]{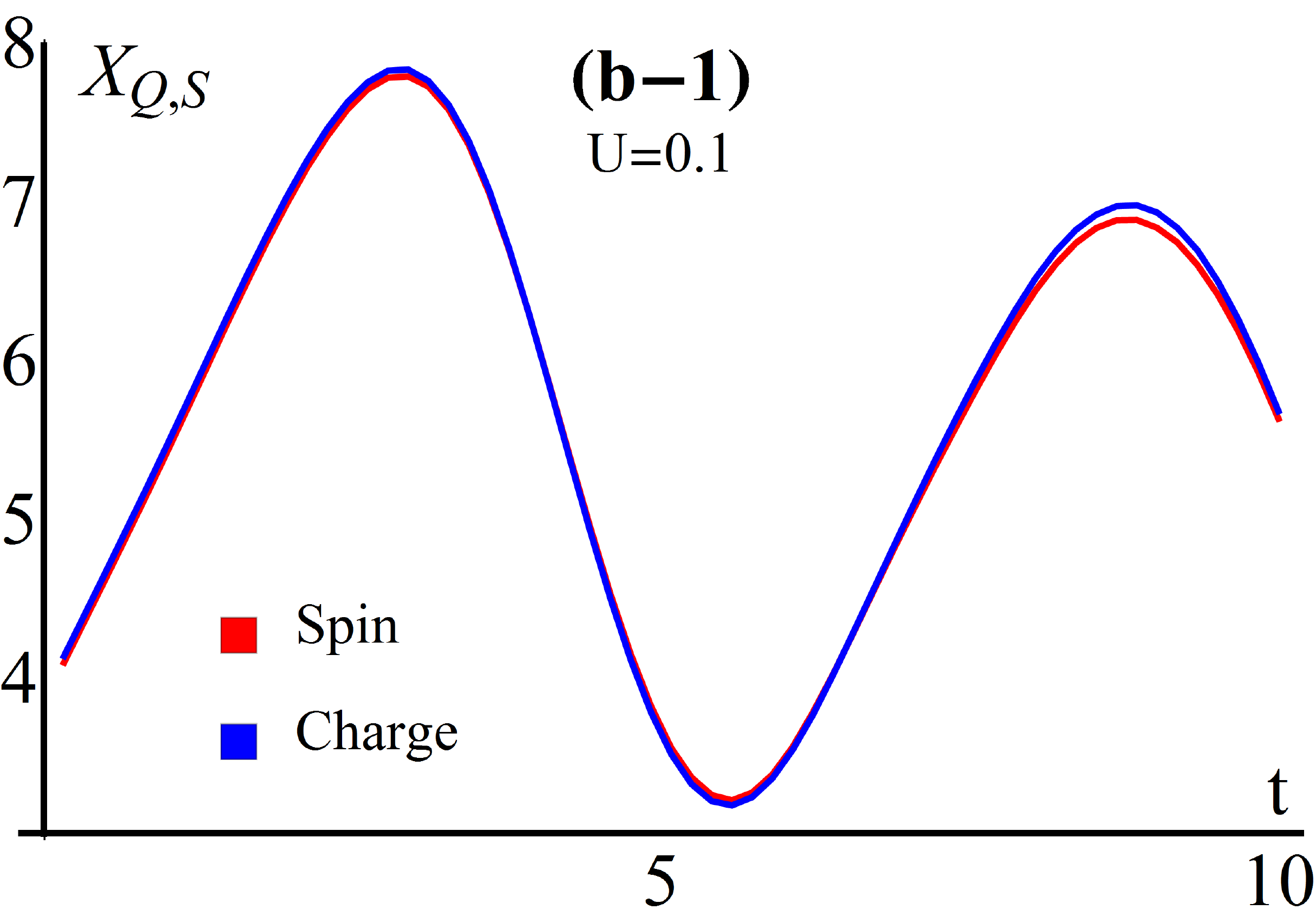}\\
  \includegraphics[width=0.23\textwidth]{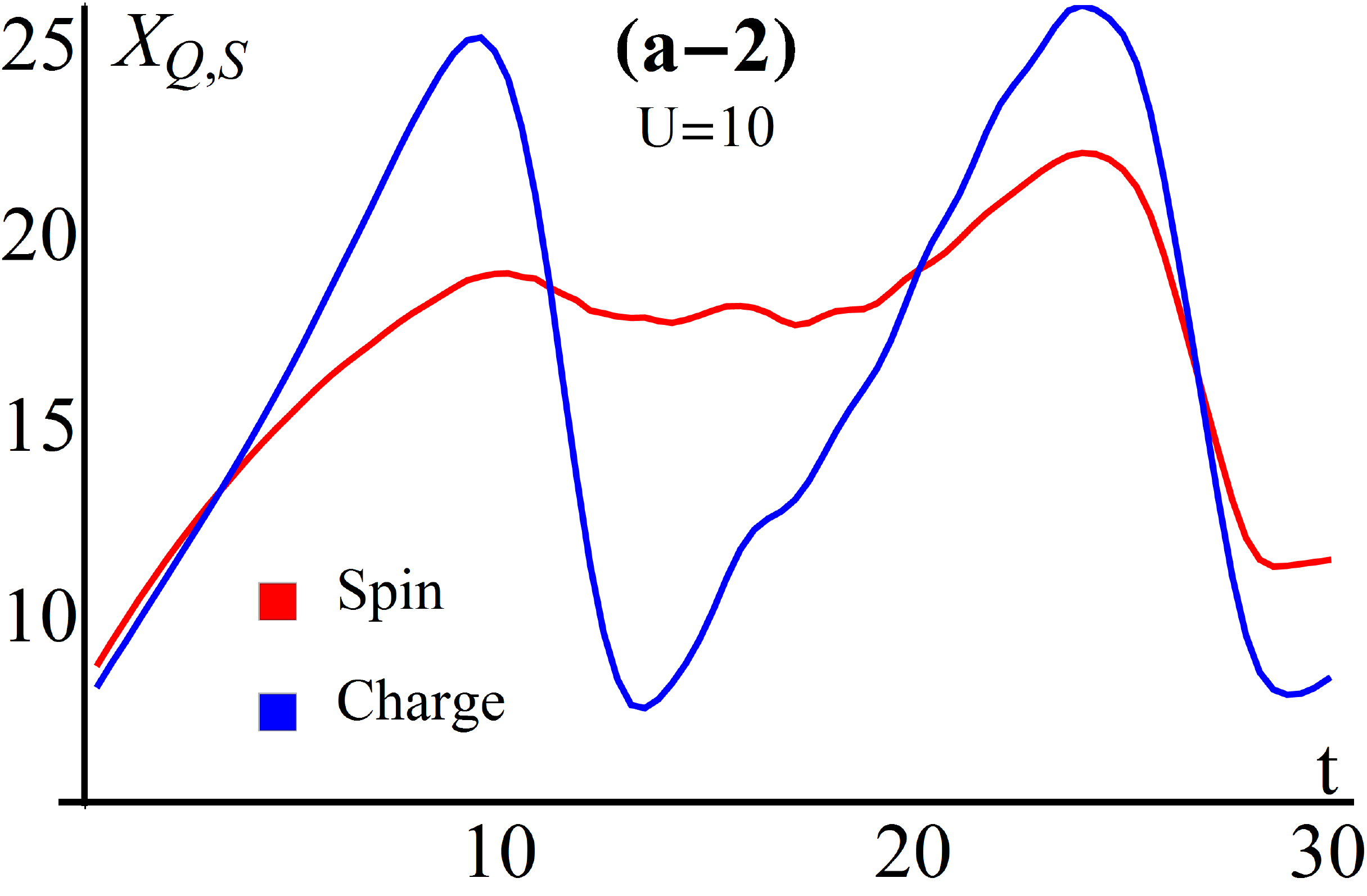}\hspace{3mm}
  \includegraphics[width=0.23\textwidth]{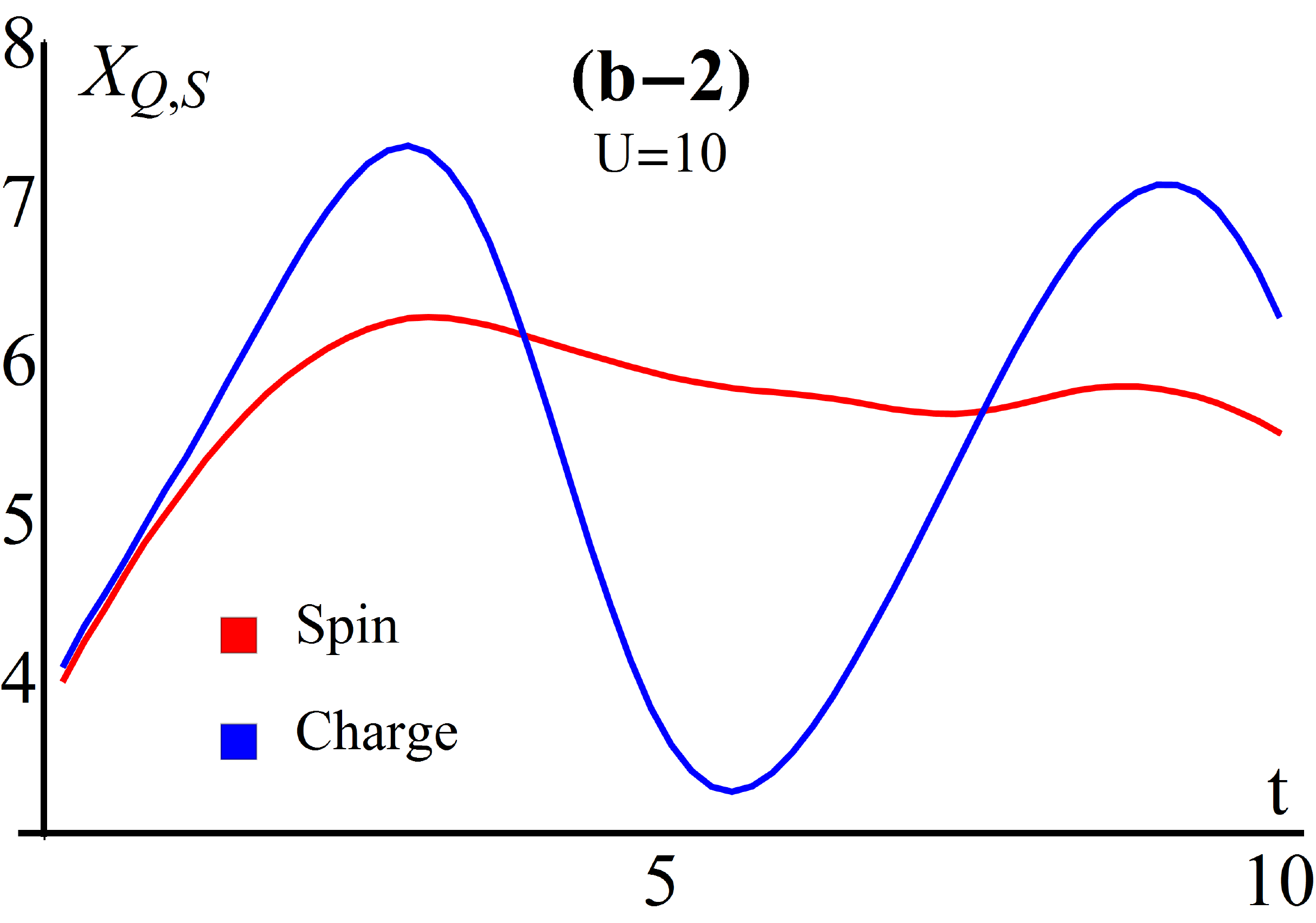}\\
  \includegraphics[width=0.23\textwidth]{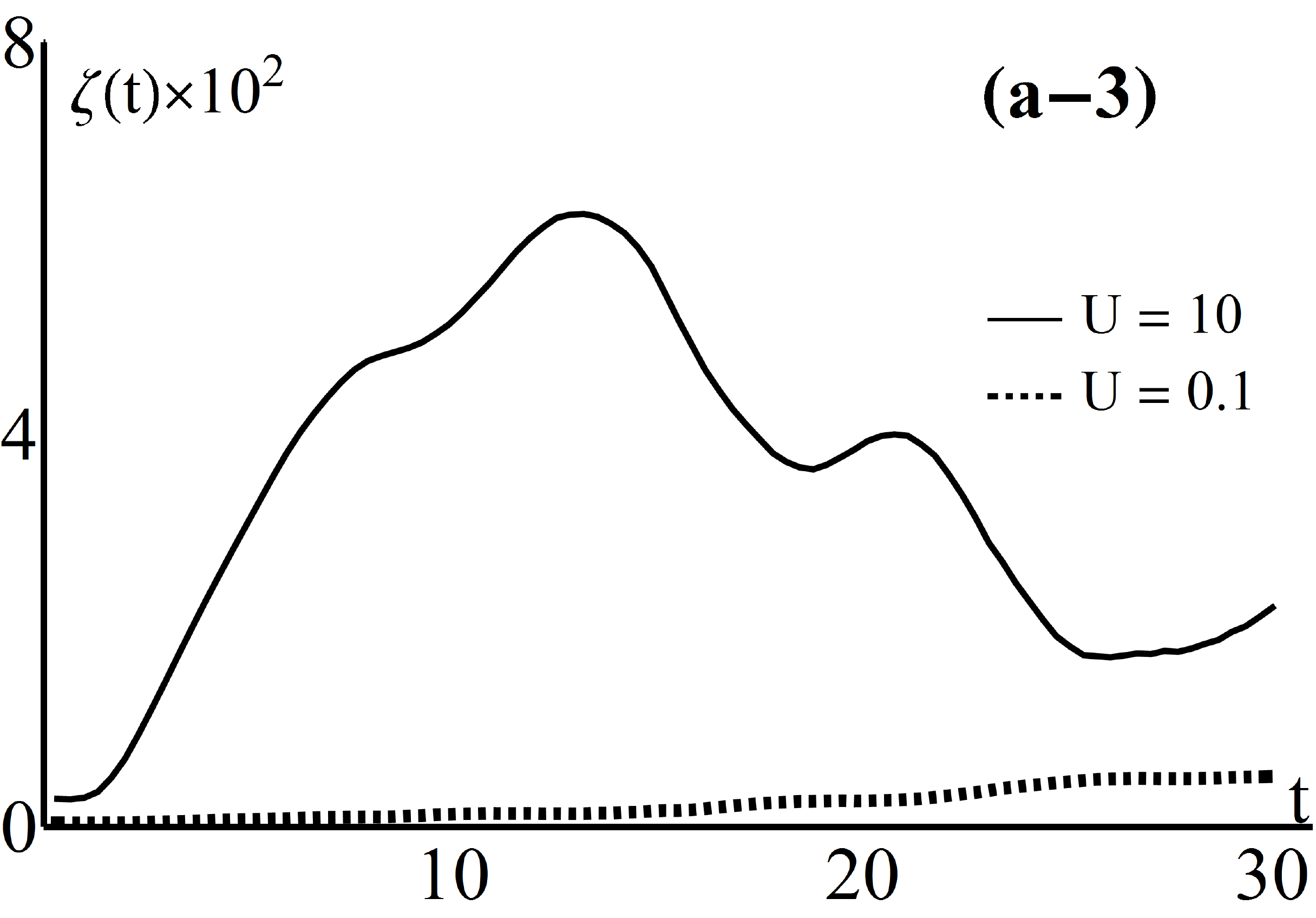} \hspace{1mm}
  \includegraphics[width=0.23\textwidth]{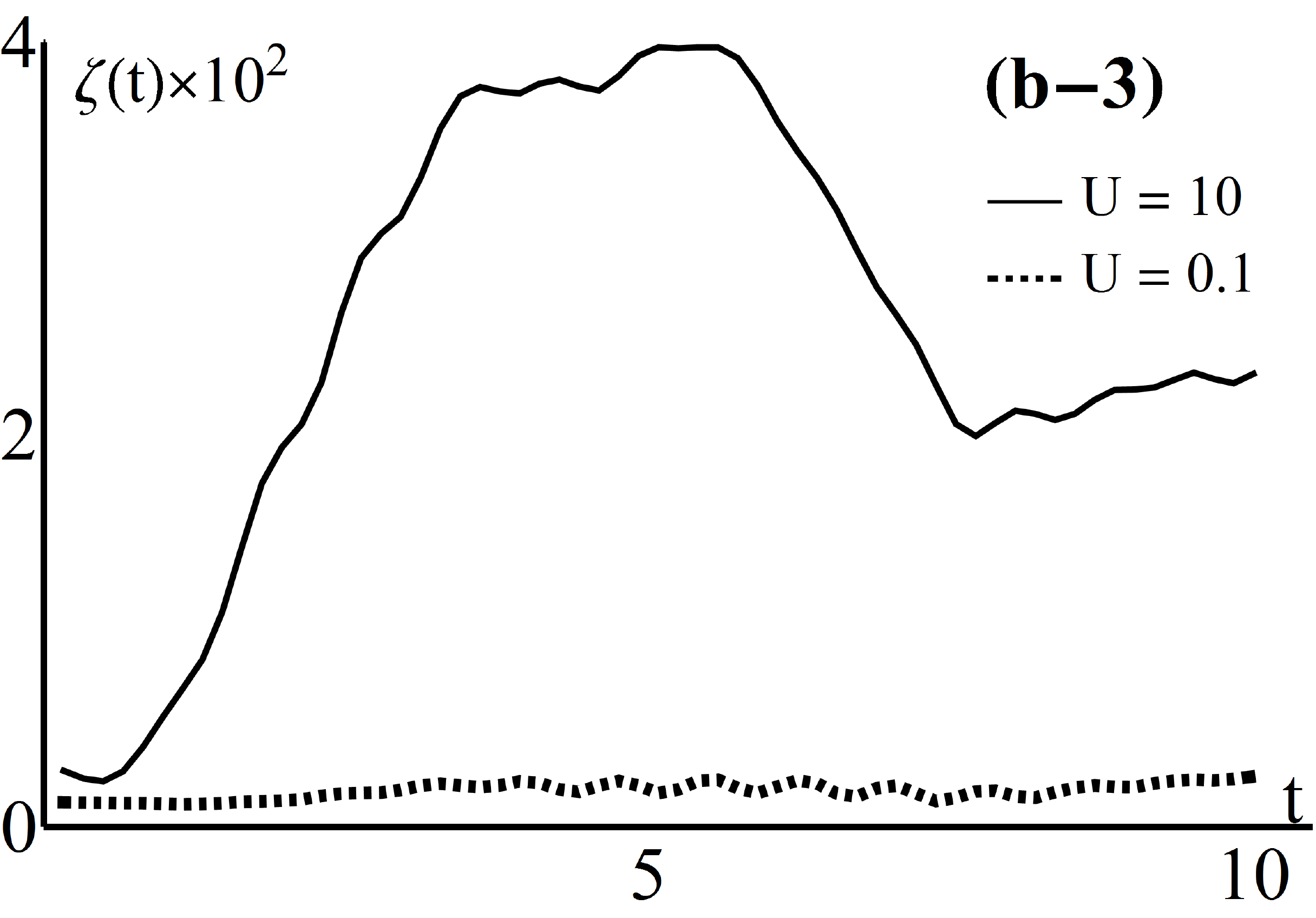}
  \caption{Worldlines of charge (blue) and spin (red), 
     $X_{Q,S}$ defined in Eq.~\eqref{xqxs.eqn} along with
     the separation quantity $\zeta(t)$. (a) First column corresponds to 1D chain, and the second
     column (b) corresponds to a two-leg ladder. Panels (a-1) and (b-1) present the 
     charge and spin wordlines
     for $U=0.1$ while (a-2) and (b-2) correspond to $U=10$. The third row represents the 
     separation quantity for chain (a-3) and ladder (b-3) for the two values of 
     $U=10,0.1$ as indicated in the legend. The one dimensional chain has $30$ sites and $N+1=3$ electrons
  and the ladder is $12\times 2$ with $N+1=5$ electrons. 
  }
     \label{1Dladder.fig}
\end{figure}

\section{Spin and charge worldlines}
In Fig.~\ref{1Dladder.fig} we have plotted the worldlines of charge and spin as defined
in Eq.\eqref{xqxs.eqn} with blue and red lines, respectively. The left column (a) 
corresponds to a 1D Hubbard chain with $L=30$ sites and $N+1=3$ electrons, 
and the right column (b) corresponds to a $12\times 2$ two-leg ladder with $N+1=5$ electrons.
Panels (a-1) and (b-1) represent the worldline of charge (blue) and spins (red) for
$U=0.1$. Panels (a-2) and (b-2) present the same data for $U=10$. 
Panels (a-3) and (b-3)
display the separation quantity $\zeta(t)$ as a function of time for the above vlaues
of Hubbard $U$ as indicated in the legend. 
Both densities are far from half-filling.
As can be seen in the first row, when the Hubbard
$U=0.1$ is small, the worldlines of spin and charge in both chain and ladder geometries 
almost coincide. For $U=10$ in the second row the worldines of spin and charge show
clear separation in both chain and ladder geometries. In both cases the charge
reaches the right end before the spin. The "returning" behaviour of the worldlines
is an artifact of periodic boundary conditions for the finite sizes
employed in this simulations as the charge (spin) density leaving e.g. the right
end, re-enters due to periodic boundary conditions from the left side which 
results in effective movement of the "center of mass" of charge (spin) to the
left. In realistic situations spin and charge keep moving in the right directions
and they will be asymptotically decoupled. Moreover one can see that
in both chain and ladder geometries spin and charge move with different
velocities, and that spin diffuses faster than charge~\cite{voit93}.

\begin{figure}[t]
  \includegraphics[width=0.23\textwidth]{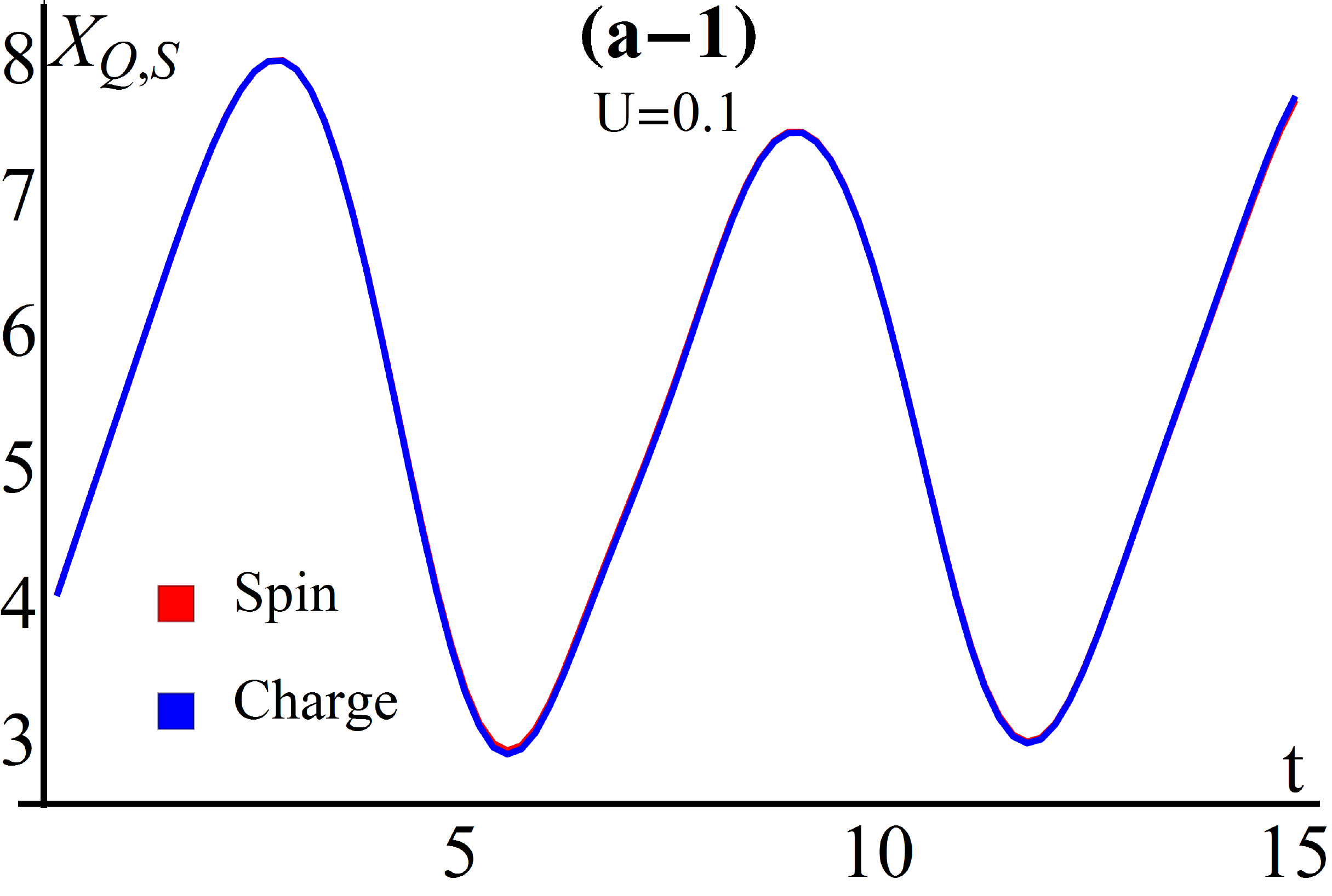}\hspace{3mm}
  \includegraphics[width=0.23\textwidth]{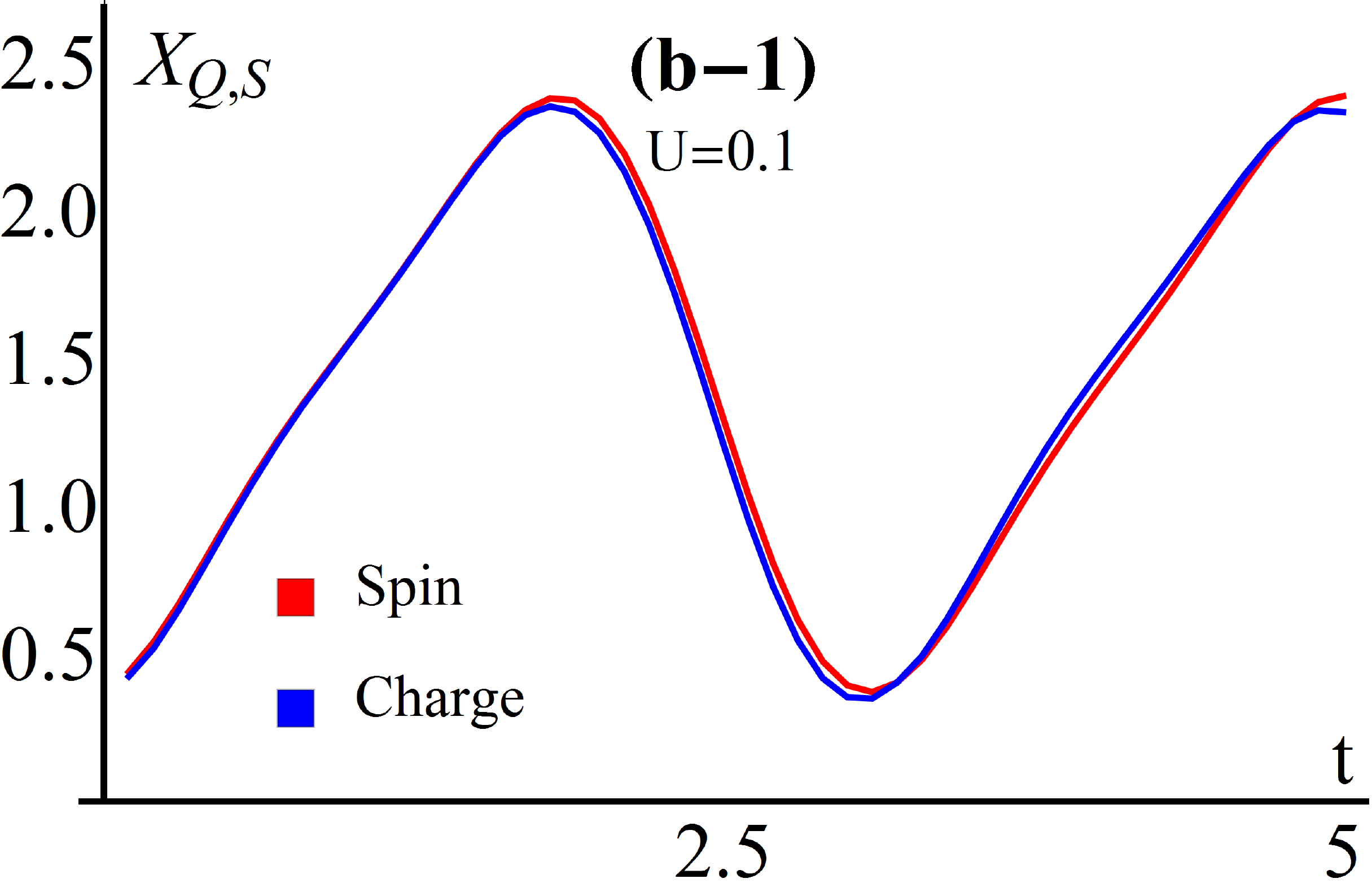}\\
  \includegraphics[width=0.23\textwidth]{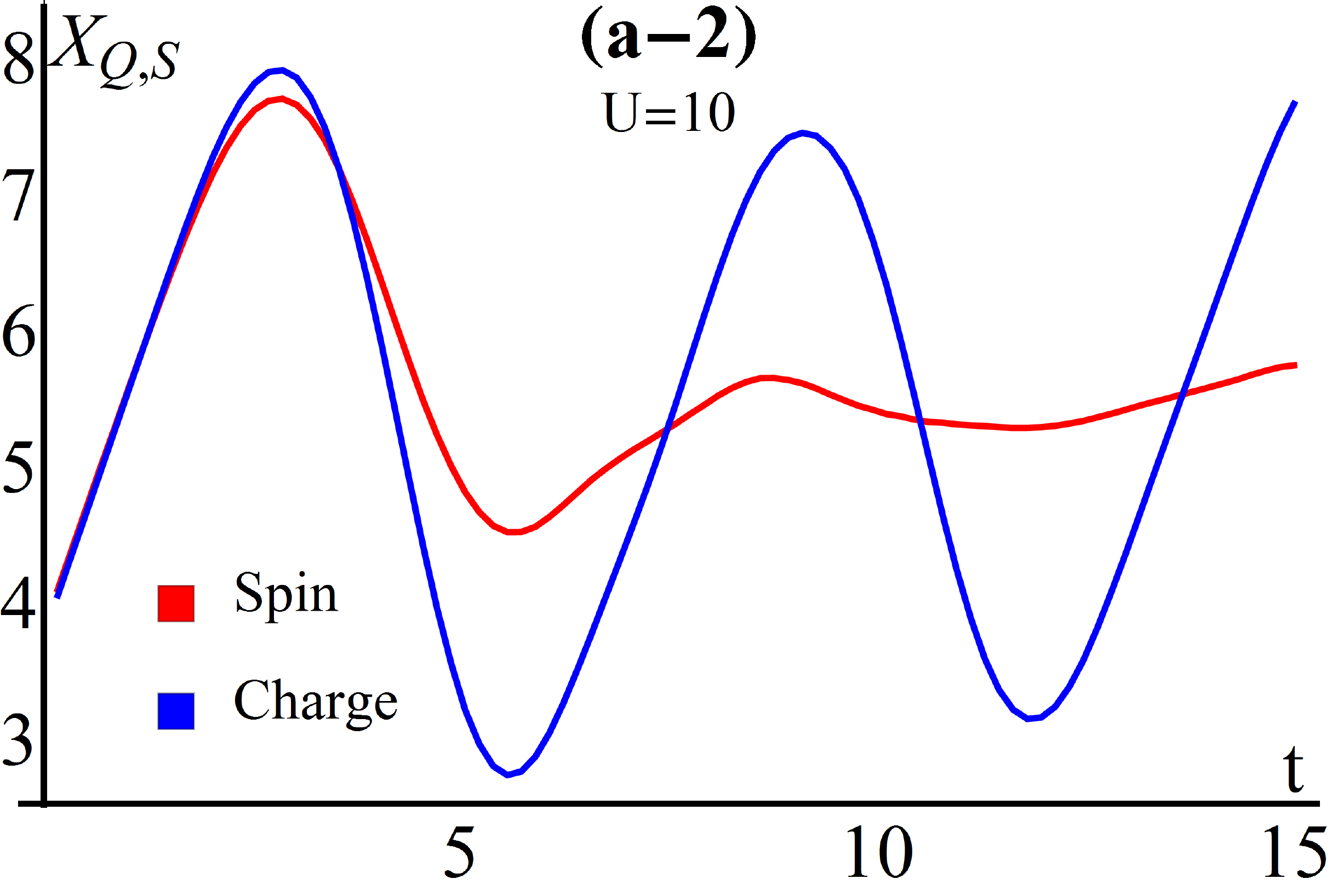}\hspace{3mm}
  \includegraphics[width=0.23\textwidth]{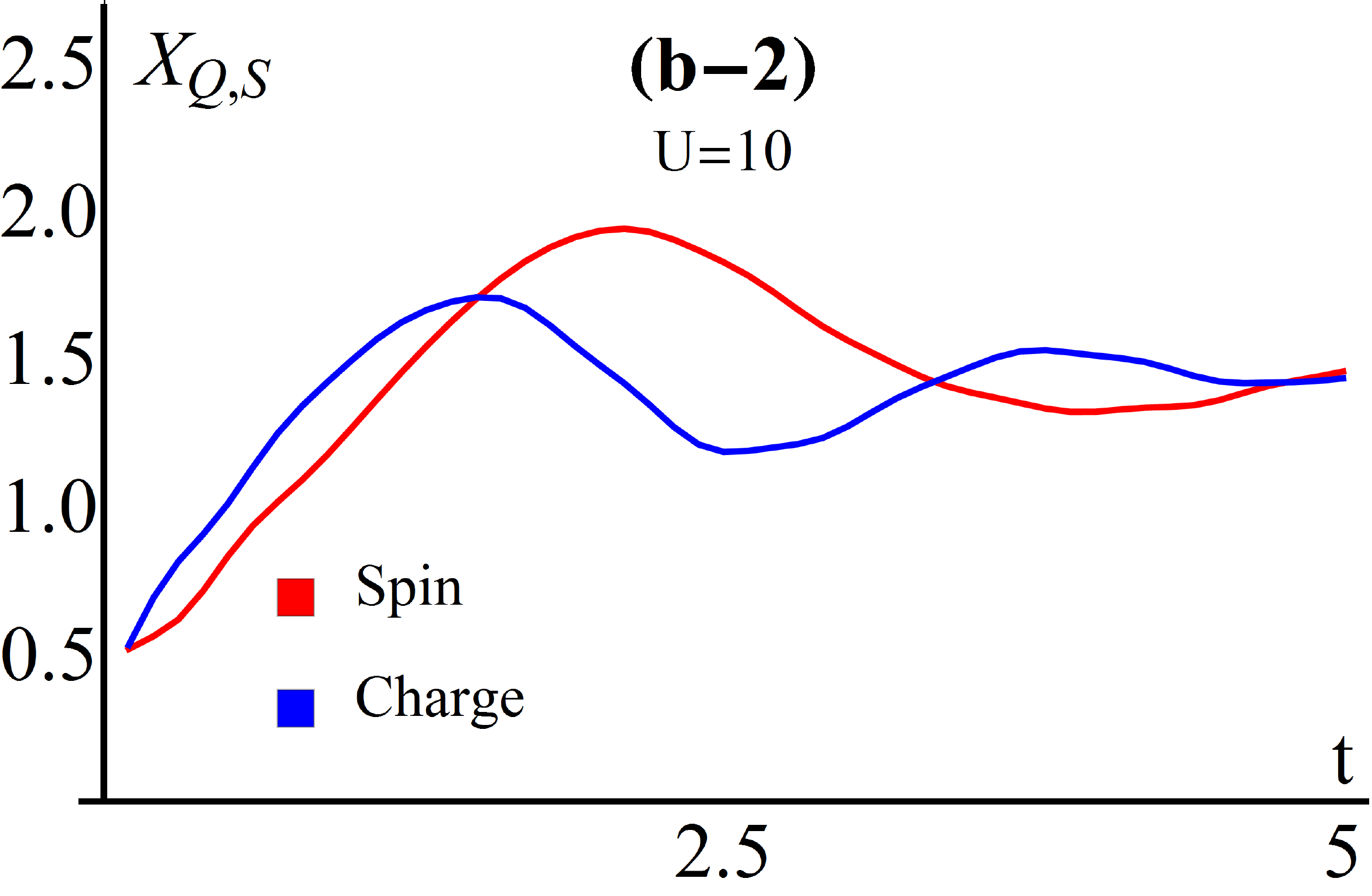}\\
  \includegraphics[width=0.23\textwidth]{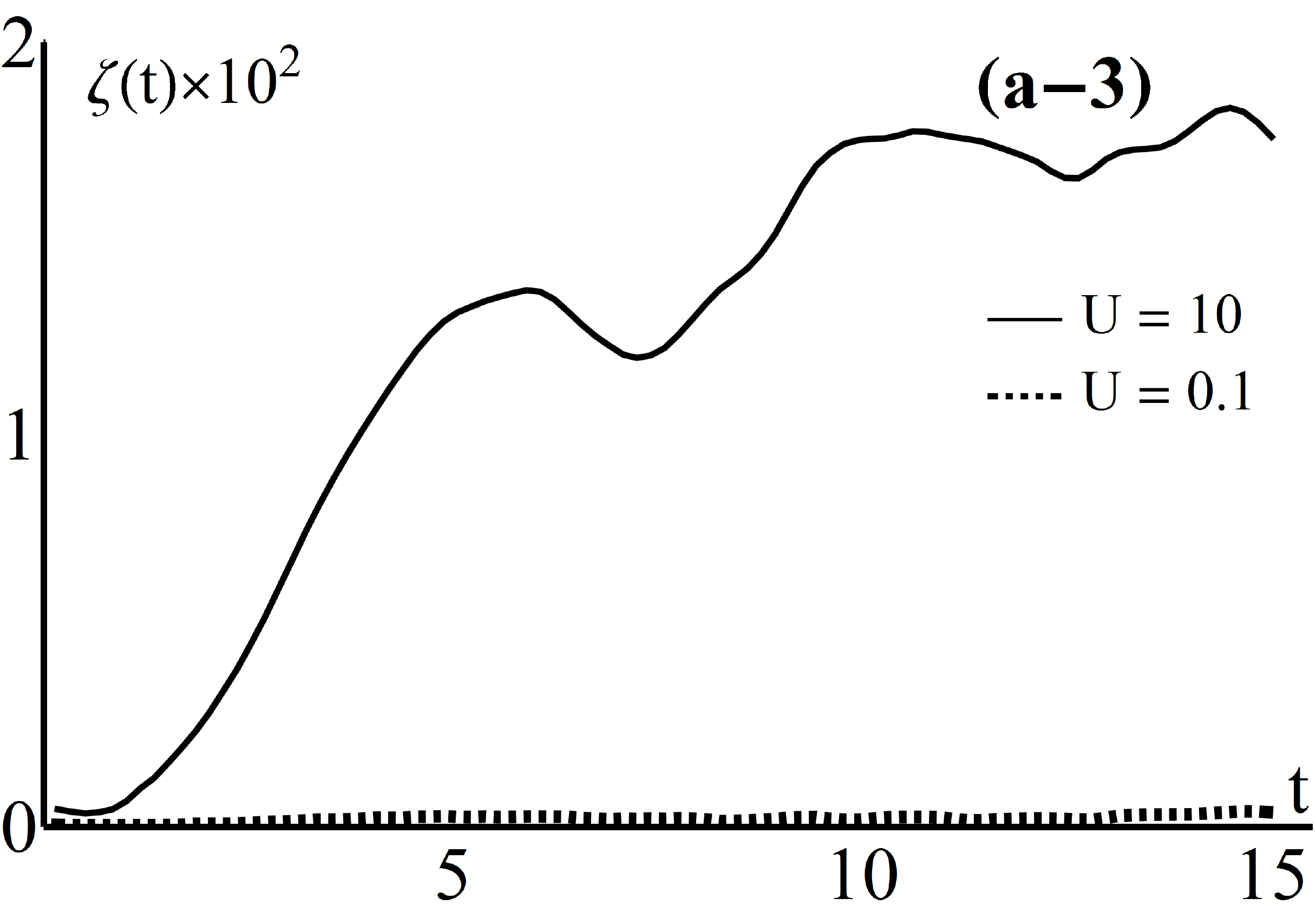} \hspace{2mm}
  \includegraphics[width=0.23\textwidth]{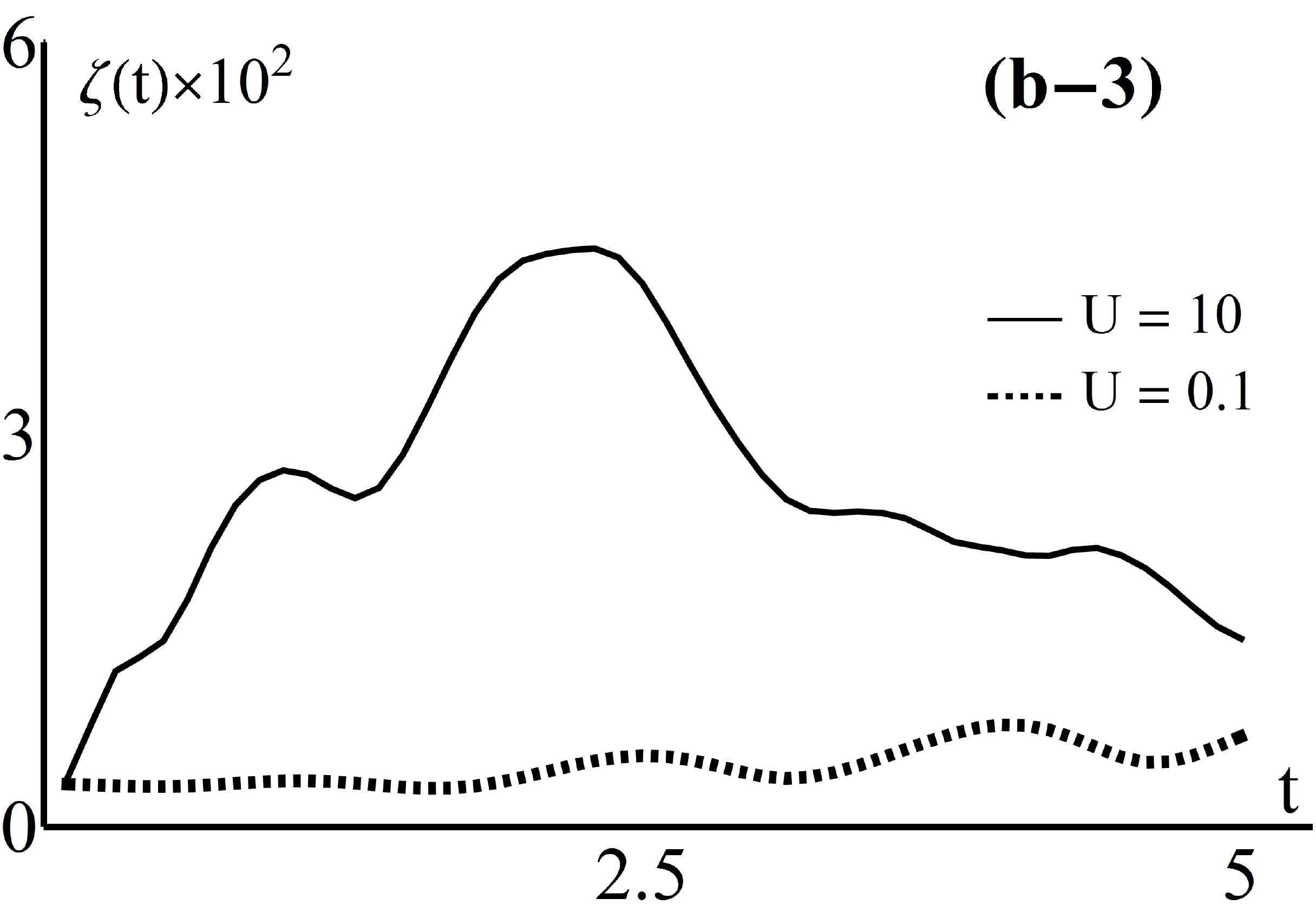}
  \caption{(Color online) Worldlines of charge (blue) and spin (red), 
     $X_{Q,S}$ defined in Eq.~\eqref{xqxs.eqn} along with
     the separation quantity $\zeta(t)$. The left column (a) corresponds to 
  $12\times 4$ lattice with $N+1=3$ electrons. The right column (b) corresponds to
  $4\times 4$ lattice with $N+1=7$ electrons. The blue (red) worldlines correspond to
  charge (spin). The first row correspons to $U=0.1$ while the second row
  corresponds to $U=10$. Third row represents the $\zeta(t)$ parameter 
  for the two lattices.}
     \label{2d.fig}
\end{figure}

Let us move to the third row of Fig.~\ref{1Dladder.fig}.
Panels (a-3) and (b-3) corresponding to chain and ladder geomtries
show the quantity $\zeta(t)$ as a function of time $t$ for two values of
Hubbard $U$ indicated in the legend. As can be seen for $U=0.1$ this 
quantity is nearly zero. By increasing $U$ to $10$, this measure of
spin-charge separation significantly deviates from zero. 
It is important to notice that the present measure of separation
in chain and ladder geometry gives comparable numbers of the order of $\sim 10^{-2}$.
As can be seen in both (a-3) and (b-3) cases the $U=10$ result
for $\zeta$ has similar behaviours: At $t=0$ it is zero by construction.
It reaches a maximum at some intermediate time scales of the order of
couple of $\hbar$ (Note that the kinetic energy scale is unit of energy). 
Then for $t\to\infty$ it again
tends to zero. 

Now that we have demonstrated the how the quantity $\zeta(t)$ works in chain and ladder
geometry, let us discuss how the quantity $\zeta(t)$ works in 2D geometry.
Fig.~\ref{2d.fig} presents the same set of data as in Fig.~\ref{1Dladder.fig}
for two 2D square lattices. The left column (a) corresponds to $12\times 4$ lattice 
with $N+1=3$ electrons and the right column (b) corresponds to $4\times 4$ lattice
with $N+1=7$ electrons, both of which are away from Mott phase at small values of Hubbard $U$. 
In both cases the charge (blue) and spin (red)
worldlines are shown for $U=0.1$ (first row) and $U=10$ (second row).
As can be seen qualitatively from the worldlines in both (a) and (b) 
lattices with different densities, the spin and charge
tend to separate for strong enough correlations. 
There is however an important difference between panels (a-2)
and (b-2) in this figure corresponding to a density of $3/48=0.0625$
and $7/16=0.4375$, respectively. If we had long range interactions, 
the role of interactions in low density case would be much more
pronounced than the higher density limit. However, since we are dealing
with the short range Hubbard interaction, the effect of Hubbard $U$
is more manifest in higher densities. That is why in panel (a-2) it takes
a longer time for the spin-charge separation to show up in $\zeta(t)$, while in panel (b-2)
at very initial time steps the velocities (slops of the worldlines) 
become different. At higher densities electric charges meet more often 
and the Hubbard $U$ will have more profound effect. 
To see this more precisely, in the third row we plot
our separation parameter $\zeta(t)$. As can be seen in both figures, 
in the $U=10$ case this parameter turns out to be on the scale of $10^{-2}$.
Therefore to the extent that spin and charge are separated in one
dimension, they show very same {\em quantitative} behaviour in 2D
and strong correlations causes spin and charge densities to propagate
with different velocities. This also demonstrates how the quantity $\zeta(t)$
encodes the difference in the worldlines of spin and charge "center of masses".
Note that densities considered here are quite below the half-filling
and the antiferromagnetic instabilities do not concern us here.

\section{Conductor versus insulator}
The behavior $\lim_{t\to\infty}\zeta(t)=0$ is indeed
an indicator of conducting properties: Since in the conducting phase both spin and charge
diffuse and ultimately spread over the whole system giving a flat 
spatial distribution of spin and charge (equal to the background value), hence $\zeta(\infty)=0$
in a conducting phase. On the
other hand, in a Mott insulating state, the spin keeps diffusing,
while the charge ultimately freezes in space giving a $\lim_{t\to\infty}\zeta(t)\ne 0$.
To this extent the long time behavior of $\zeta(t)$ -- ideally the $t\to\infty$ limit, 
but practically a couple of $\hbar/$hopping -- can serve as a 
possible "order parameter" for a Mott state. 
Let us see how does it work in various geometreis.
\begin{figure}[t]
  \includegraphics[width=0.40\textwidth]{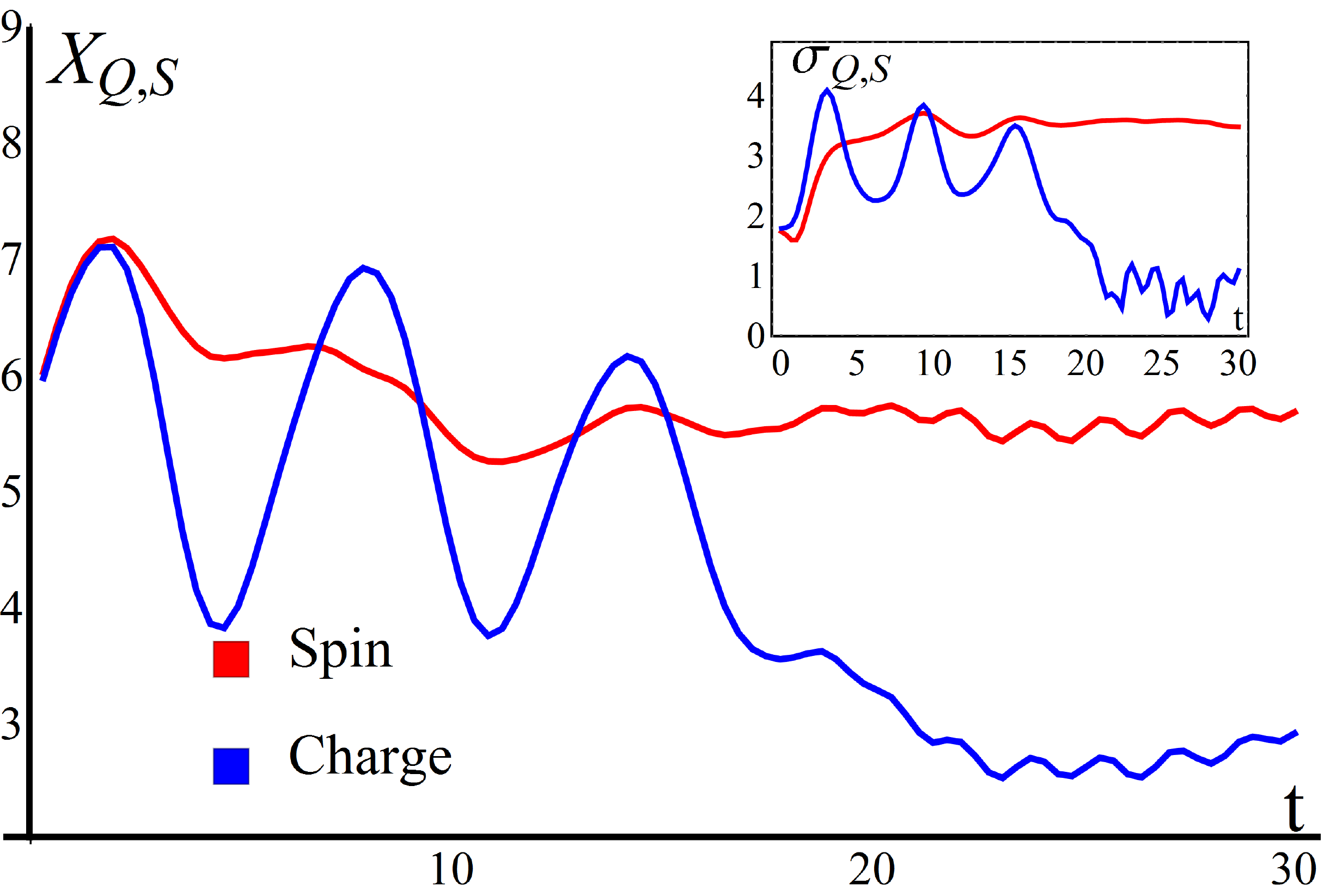}
   \caption{ (Color online) 
      The spin (red) and charge (blue) worldlines for a 
      Mott insulator at $U=30$, on a $12\times 2$ ladder with $N+1=5$ electrons.
      The inset represents the the standard deviation of the charge (blue) and
      spin (red) density. 
   }
   \label{mott-x.fig}
\end{figure}

In Fig.~\ref{mott-x.fig} we plotted the
worldlines of spin and charge for a $12\times 2$ ladder with $5$ electrons
and $U=30$. The inset of the 
top panel shows the spread of the center of mass of charge and spin.
The worldline of the spin shows that the spin center of mass saturates
towards the center of ladder which due to periodic boundary condition 
means that the spin has uniformly spread all over the lattice.
To see this more clearly, in the inset we have shown the standard deviation 
of the center of mass of charge and spin.
The charge woldline on the other hand shows that the charge freezes at a 
different point of the lattice. At initial times steps the charge density 
moves along the ladder. But beyond a certain time scale the charge density
starts to feel the effect of strong Hubbard $U$. This onset time is controlled
by Hubbard $U$ and the density of electrons. Once the charge realizes that
it lives in a Mott insulator, it freezes at some point.
The inset of the top panel shows the spread of charge (blue) and spin (red)
densities. As can be seen the spin has undergone a diffusion and has been delocalized
all over the ladder, while the charge has been localized in some spot of the ladder.
The initial three peak structures in both worldines and spread of the
charge density indicate that due to periodic boundary condition the 
charge density has revolved three times across the ladder length. 

The localization of charge and delocalization of spin density is a
distinct feature of Mott insulating state which has been naturally
captured in the worldlines of spin and charge densities. What our 
worldlines indicate further is that for the charge added to a Mott
state it takes some time to "figure out" that will live in a Mott
insulating environment. Before this threshhold time, $t_{\rm tr}$ the charge density
keeps moving until the charge density learns enough about the
dynamics of the Mott insulating Hamiltonian after which it stops moving. 
By that time the spin density has already delocalized all over the
lattice. This threshhold time by uncertainity principle would correspond 
to a threshchold energy, meaning that a Mott insulator would conduct
above a certain energy scale (voltage) $V_{\rm tr}\sim \hbar/ t_{\rm tr}$
that can be interpreted as the breakdown of Mott insulating phase~\cite{Oka}.

\begin{figure}[t]
  \includegraphics[width=0.45\textwidth]{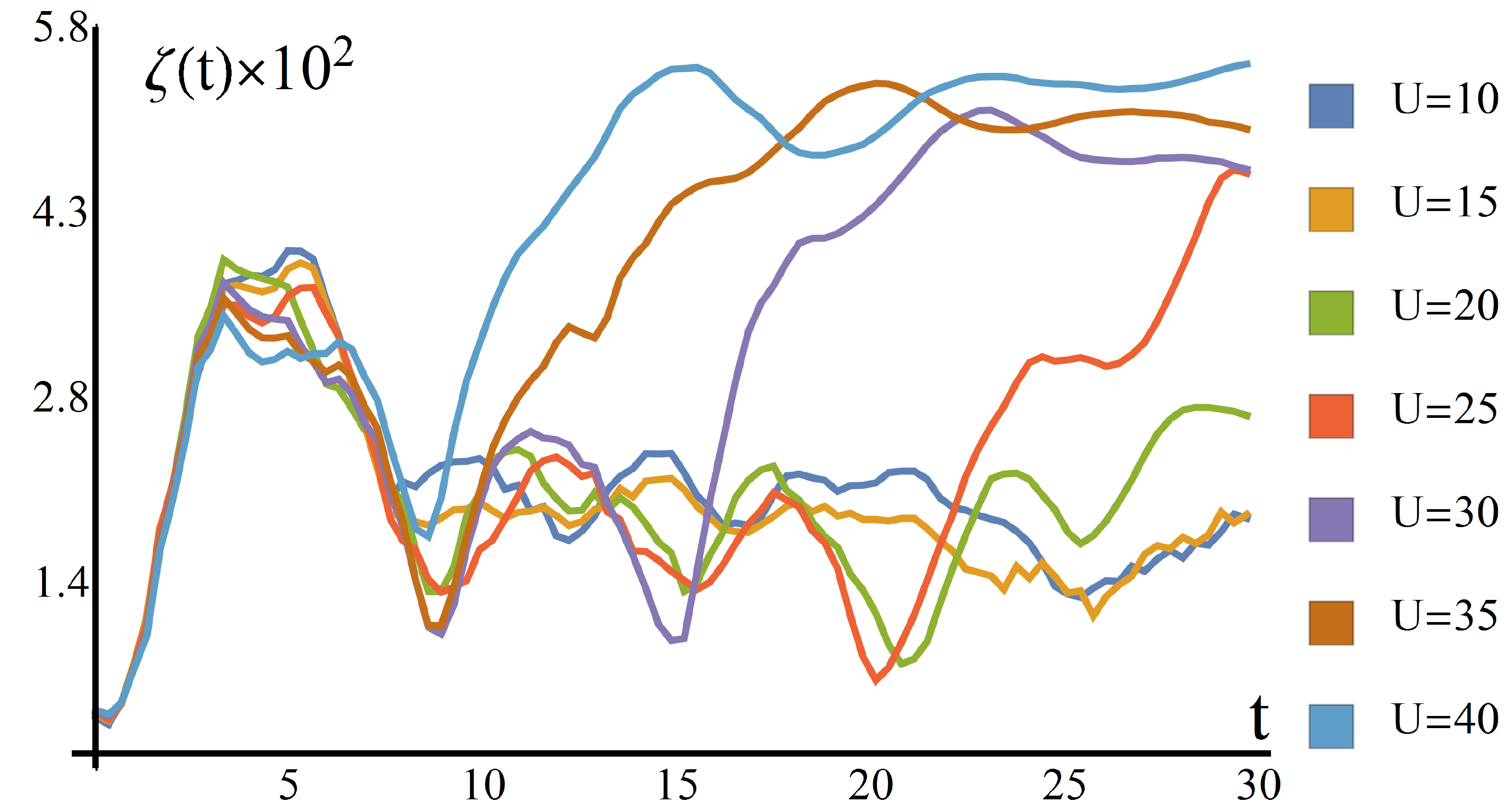}
   \caption{ (Color online) The $\zeta(t)$ for $12\times 2$ ladder 
      with a fixed $N+1=5$ electrons and variable Hubbard $U$ values
      indicated in the legend. 
   }
   \label{mott-zeta.fig}
\end{figure}

Now we are ready to demonstrate how the Mott transition shows up in our 
$\zeta(t)$ profiles. For this purpose in Fig.~\ref{mott-zeta.fig} 
we have plotted this function for the $12\times 2$ ladder with a fixed
$N+1=5$ number of electrons for various values of Hubbard $U$ parameter. 
The early time behaviour of the $\zeta(t)$ is quite similar for all
values of $U$ signalling that it is a transient behaviour. Beyond a certain
time scale to the left of $t=10$ the separation quantity starts to 
know about the differences in Hubbard $U$. For smaller values of $U$
the $\zeta(t)$ decreases at longer times. Let us argue that this 
corresponds to a conducting state: The spin density already spreads
and delocalizes itself irrespective of whether it is in the Mott
state or conducting state. However for smaller values of $U$ corresponding
to conducting state, the charge density diffuses as well, and will
eventually spread uniformly over the entire ladder giving $\zeta(t\to\infty)=0$.
Therefore the decreasing behaviour of $\zeta(t)$ at large $t$ is characteristic
of a conducting state. In the Mott state in contrary the charge freezes 
at some location after a threshhold time $t_{\rm tr}$. This prevents the 
$\zeta$ function from diminishing. Therefore in the Mott state $\zeta$ is
expected to saturate to a non-zero value. That is why for values of $U$ larger
than $25$ the $\zeta$ function takes up in long times and saturates to a 
non-zero value. 

\begin{figure}[b]
\subfloat[]{\includegraphics[width = 3in]{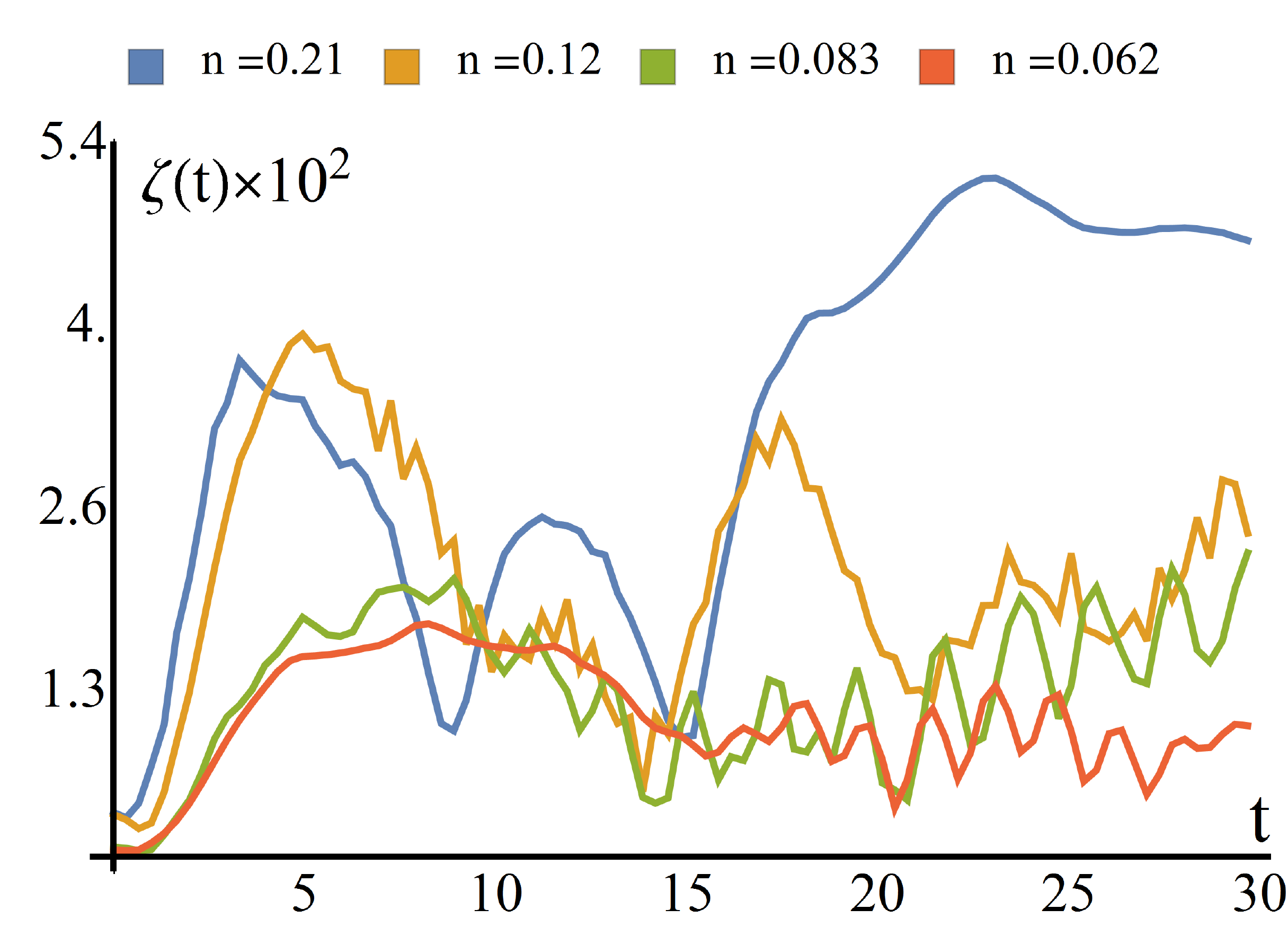}}\\
\caption{ (Color online) Temporal profile of $\zeta(t)$ for a fixed
value of $U=30$ and various values of filling fraction $n=(N+1)/L$ with
$N+1=5,L=12\times 2$ $(n=0.21)$, $N+1=3,L=12\times 2$ $(n=0.12)$, 
$N+1=3,L=12\times 3$ $(n=0.083)$ and $N+1=3,L=12\times 4$ $(n=0.062)$.
}
\label{mott-n.fig}
\end{figure}

While Fig.~\ref{mott-zeta.fig} indicates that the temporal profile of  $\zeta(t)$ 
contains information about the charge localization in the Mott state for
various values of $U$, in Fig.~\ref{mott-n.fig} for a fixed value of $U=30$ 
we present the temporal profile of $\zeta$ for different values of filling factor $n$
indicated in the figure. In this figure we have used various lattice sizes indicated
in the figure. For $U=30$, only at the highest filling factor $n=0.21$ reported here
we find a saturation behaviour in $\zeta(t)$ indicating a Mott insulating phase for 
this filling. For lower fillings, even a value of $U$ as large as $30$ is not able 
to give rise charge localization behaviour. For a fixed $U$ when the filling fraction
is larger, the probability of two electrons to come across each other at the same site
(the double occupancy) increases. Therefore at larger $n$ a given value of Hubbard $U$
does a better job at suppressing the double occupancy. That is why in this figure
for lower filling fractions the $U=30$ value is not able to localize the charges
and $\zeta(t)$ decreases at large $t$.

\section{Summary and conclusion}
In this work we have introduced a quantity based on the difference
in the fluctuations of the spin and charge around their center of mass
whose temporal behaviour and magnitude contains information about the 
spin-charge separation. Within the present $\zeta(t)$ function we find that
the spin and charge do separate in 2D to the extent that they do in 1D. 
Further we found that the long time behaviour of this function differs
for conducting states and Mott insulating states. It is zero for conducting
states and non-zero for Mott insulating states. It can therefore serve
as an order parameter for the Mott state. Furthermore, a threshhold time
scale in the behaviour of $\zeta(t)$ that separates transient non-equilibrium
behaviour from the long-time equilibrated situation 
reveals information about the voltage at which a Mott insulating 
phase breaks down. 

Indeed a a two-dimensional version of Luttinger lequid theory has been
postulated by Anderson~\cite{AndersonHall} and has been used by him 
to explain the temperature dependence of the Hall effect
in normal state of cuprate superconductors~\cite{Anderson2DLuttinger}.
The present work on quantification of the spin-charge separation demonstrates
that spin and charge excitations of the two dimensional Hubbard model are
separated and therefore the ground state of the 2D Hubbard model in low-doping
does not appear to be a Fermi liquid.
Further research is needed to understand the properties of 
such a non-Fermi liquid state in two-dimension.

\section{Acknowledgement}
SAJ was supported by Sharif university of technology and Alexander von Humboldt
foundation.

\bibliographystyle{apsrev}

\end{document}